\documentclass{epl}
\usepackage{amsmath}
\title{A Tight-Binding Investigation of the \chem{Na_xCoO_2} Fermi Surface 
}
\shorttitle{\chem{Na_xCoO_2} Fermi surfaces}

\author{M. D. Johannes, D. A. Papaconstantopoulos, D. J. Singh, M. J. Mehl}
\institute{Code 6391, Naval Research Laboratory, Washington, 
D.C. 20375}


\begin{document}
\maketitle

\begin{abstract}
        We perform an orthogonal basis tight binding fit to an LAPW calculation of paramagnetic Na$_x$CoO$_2$ for several dopings.  
The optimal position of the apical oxygen at each doping is resolved, revealing a non-trivial dependence of the band structure and
Fermi surface on oxygen height.  We find that the small e$_{g'}$ hole pockets are preserved throughout all investigated dopings
and discuss some possible reasons for the lack of experimental evidence for these Fermi sheets.  \end{abstract}

The compound Na$_x$CoO$_2$ has been synthesized for a wide range of Na contents, 0.3 $<$ x $<$ 0.9, and exhibits a
variety of unusual phenomena that are strongly dependent on doping level.  Study of Na$_{0.5}$CoO$_2$ for its
surprisingly good thermoelectric properties began several years ago \cite{ITYS97} and more recently, attention has
been focused on the hydrated compound Na$_{0.35}$CoO$_2$$ \cdot$1.3H$_2$O which undergoes a superconducting phase
transition at T $\sim$ 4.5K \cite{KTHS+03, RES+03}.
	
The behavior and characteristics of Na$_x$CoO$_2$ do not vary smoothly along the spectrum of Na contents, but instead split very generally
into two separate metallic phases separated at $x$=0.5 by an insulating phase \cite{QHMLF+04}.  The low Na content region at $x <$0.5 is
characterized by conventional metallic behavior \cite{MLF+04} and a nearly temperature independent susceptibility \cite{RJBCS+03,FCC+03}
indicating a Pauli paramagnetic precursor to the superconducting state.  The linear specific heat coefficient $\gamma$ is measured to be
$\sim$ 12-16 $\frac{mJ}{mol K^2}$ which shows only light mass renormalization \cite{RJBCS+03,BGU04}. In contrast, the high Na region at $x >$
0.5, though also metallic, exhibits a Curie-Weiss-like susceptibility \cite{BCS+04,TMRU+03,JSJHB+03}.  A higher (compared to the previous
metallic phase) measured $\gamma$ of 25-30 $\frac{mJ}{mol K^2}$ indicates that mass renormalization in this region is substantial
\cite{MBBB+03,JSJHB+03,TMRU+03}.  Interestingly, the Kadawaki-Woods ratio, a comparison of $\gamma$ to the quadratic coefficient of
resisitivity, which is often used as a measure of electron-electron interaction, is reported to be the largest ever measured and considerably
field dependent \cite{SYL+04}.  This may evidence a nearby quantum critical point and indeed, $\mu$sR and transport measurements show a
magnetic transition in single crystals at T=22K for dopings of $x$= 0.7 and $x$= 0.75, usually interpreted as a spin density wave
\cite{BCS+04,JSHI+03,DPATB03}.  Strong ferromagnetic spin fluctuations have been predicted based on LDA calculations \cite{DJS02} and have
been detected by neutron scattering \cite{ATB+03} when $x$= 0.7, but with slightly higher electron count at $x$=0.85, a bulk
anti-ferromagnetic transition has been reported \cite{SBCB+03}, suggesting the presence of more than one type of magnetic interaction. 
The insulating phase is very narrowly centered at $x$=0.5 and recent reports show both charge and magnetic order \cite{QHMLF+04},
though earlier investigations, most notably work by Terasaki et al \cite{ITYS97,YANM+99,IT02}, found this system to be metallic.  
Measurements of $\gamma$ $\sim$ 40-56 $\frac{mJ}{mol K^2}$ are higher than for any other doping \cite{YANM+99,IT02} and NMR measurements
indicate Curie-Weiss behavior and charge separation \cite{RRAG99}.

Here we present a highly accurate tight-binding parameterization of Na$_x$CoO$_2$, obtained for a range of apical oxygen
heights at each of three dopings: $x$=0.3, 0.5, and 0.7, meant to be representative of the three general phases.  We
anticipate that this Hamiltonian will be valuable as accurate, material-specific input to model systems that can incorporate
the important effects of correlation.  We point out that, since the system is multi-band in nature, a strictly
two-dimensional triangular lattice model is insufficient to fully account for all dispersions.  In contrast to earlier
tight-binding models \cite{WKSM03b,BKBSS03}, we include both interlayer couplings and all five Co-$d$ orbitals as a basis.  
The nearly perfect reproduction of LDA Fermi level crossings within our tight-binding scheme allows a detailed and accurate
examination of doping and structural Fermi surface dependencies. We show that the band structure is strongly coupled to the
apical oxygen height which itself depends on the amount of charge donated to the CoO$_2$ plane by Na ions.  A rigid band
picture is insufficient to describe the evolution from small $x$ where the band structure is nearly two dimensional to large
$x$ where non-negligible inter-planar coupling causes heavy distortion of one Fermi sheet.  An important effect of increased
three-dimensionality in the high Na compounds is the preservation of the small hole pockets at the K points of the Brillouin
zone (BZ).  We discuss possible reasons that these pockets are absent from ARPES measurements \cite{MZH+03,H-Y03}.

Through a series of calculations with an LAPW implementation (Wien2k) \cite{Wien2k} of DFT formalism, we varied the Na level, $x$, of
Na$_x$CoO$_2$, holding the total volume constant at 509.553 a.u.$^3$.  Though important effects may derive from the specific location of
Na ions, particularly in the insulating compound at $x$=0.5, where Na and charge order are thought to cooperate, we treat Na in the
virtual crystal approximation (VCA) and concentrate only on effects stemming directly from variation of electron number.  Specifically,
we treat every Na site as fully occupied, but with a fictitious ion of atomic charge 10.$x$, rather than a true Na atom of charge 11.
This approximation is most valid when the Na states are far from the Fermi energy, as they are Na$_x$CoO$_2$ for all $x$, and in fact, 
comparison of VCA and Na ordered calculations show little difference \cite{MDJ04}.  The apical
oxygen height was relaxed for a number of dopings and shown to increase non-linearly with Na addition.  The optimal O position as a
fraction of the lattice constant (a= 5.366824 a.u.) for $x$ = 0.3, 0.5 and 0.7 was found to be 0.0811, 0.0834, and 0.0864 respectively.  
The six Co-derived bands near the Fermi energy in each of these first-principles calculations form the basis for our tight-binding fit.

We fit the LAPW eigenvalues using the NRL tight-binding method \cite{DAP03}.  Assuming that the important features of the band
structure are due to Co and O ion interactions, we used a basis set consisting of 5 Co-d orbitals for each of two Co sites and 3 O-p
orbitals for each of four O sites resulting in 25 Slater Koster (SK) hopping parameters and 5 on-site energies to vary.  The six bands
(4 e$_{g'}$ and 2 a$_{1g}$) nearest the Fermi energy are cleanly separated from the e$_g$ bands above and O 2p bands below by an energy
gap \cite{DJS00} allowing them to be isolated from all other bands during the fit by a weighting of the eigenvalues.  
The orthogonality of our basis set was enforced and site symmetry taken into account by requiring that on-site energies of
orbitals in a given irreducible representation be the same.  We kept neighbors up to a distance greater than half the c-axis lattice
parameter to account for inter-planar coupling.  Because this basis set and number of neighbors results in a large
amount of freedom, the parameters that fit the LAPW band structure accurately are not unique, and several solutions which nicely
reproduced the calculated dispersions were found.  In order to distinguish the desired physically relevant solution ({\it i.e.}
atomic-like orbital picture), we compared the tight-binding eigenvectors to the orbitally resolved LAPW band structure at each
k-point.  The criteria that the characters of each eigenvalue must qualitatively match was used to identify a single set of parameters 
as the correct one.

\begin{figure}
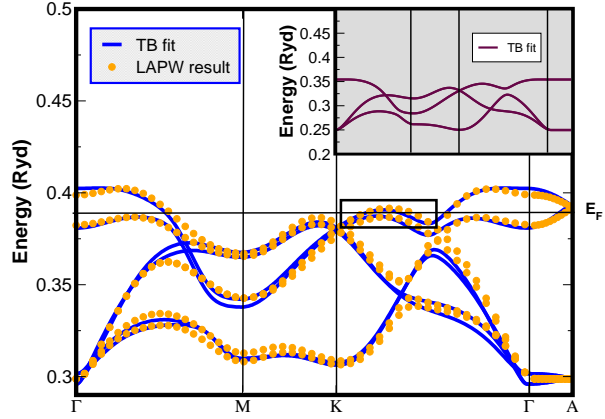

\onefigure[width = 3.1 in]{1.eps}
\caption{A
tight-binding fit to the $\delta$ = 0.7 optimal apical oxygen height band structure.  The rectangular box
indicates the Fermi level crossing responsible for the e$_{g'}$ elliptical hole pockets.  The upper inset
has an expanded c-axis to eliminate inter-planar coupling.}
\label{fit} \end{figure}

        For each doping, we fit to band structures with three different apical O heights: one above, one below and one at the optimal
position.  Since the NRL-TB parameters are expressed as polynomials in distance, we could fit all O positions at each doping
simultaneously, thereby explicitly including O height dependence in the parameterization.  The familiar SK elements are then easily
obtained for any structure by evaluating the NRL-TB parameters at a specific distance.  A table of SK elements and on-site energies
for all three dopings is provided in Tables \ref{SK} and \ref{onsite}, with the O position set at its relaxed height in each case. A
typical fit (that of $x$=0.7) is shown in fig. \ref{fit}.  The Fermi level crossings are reproduced almost exactly, though there are
small discrepancies in dispersion lower in energy, most notably between $\Gamma$ and M.  The rectangular box highlights the e$_{g'}$
band as it breaks above the Fermi energy; this feature is responsible for the small hole Fermi sheets which will be the central
concern of this paper.

\begin{table}
\caption{Hopping Integrals (meV)}
\begin{largetabular}{lcccccccr}
\hline \multicolumn{2}{c}{} & \multicolumn{2}{c}{Co-O} & \multicolumn{2}{c}{O-O} &
\multicolumn{3}{c}{Co-Co} \\ \cline{3-4} \cline{5-6} \cline{7-9}
\multicolumn{2}{c}{} & $dp\sigma$ & $dp\pi$ & $pp\sigma$ & $pp\pi$ & $dd\sigma$ & $dd\pi$ & $dd\delta$ \\ \hline
& \multicolumn{1}{c}{1st n.n.}  & -1286  & 1054 & 112 & 30 & -326  & 34  & 34
\\
 & \multicolumn{1}{c}{2nd n.n.}  & -33  & 21 & 78 & 228 & 2  & 72  & -68  \\
d=0.3 & \multicolumn{1}{c}{3rd n.n.}   & 0 & -8 & 56  & 132  & 0  & 13  & -62   \\ 
& \multicolumn{1}{c}{4th n.n.}  & & & 53 & 122 & & & \\
& \multicolumn{1}{c}{5th n.n.}  & & & 18 & 30 & & & \\ 
& \multicolumn{1}{c}{1st n.n.}  & -1422  & 1003 & 250 & 20 & -405 & 63 & 36 \\
& \multicolumn{1}{c}{2nd n.n.}  & -53  & 33 & 135 & 198  & 33 & 63   & -83 \\
d=0.5& \multicolumn{1}{c}{3rd n.n.}  & -2  & -3  & 107 & 133 & 4 & 11 & -57  \\ 
& \multicolumn{1}{c}{4th n.n.}  & & & 100 & 115 & & & \\ 
& \multicolumn{1}{c}{5th n.n.}  & & & 44 & 27 & & & \\ 

& \multicolumn{1}{c}{1st n.n.}  & -1551  & 929 & 295 & 59  & -433 & 90 & 16  \\ 
& \multicolumn{1}{c}{2nd n.n.}  &-88  & 64 & 309 & 187 & 52 & 56 & -97  \\
d=0.7& \multicolumn{1}{c}{3rd n.n.}  &-4   &-9  & 204 & 156 & 6 &10  & -47 \\ 
& \multicolumn{1}{c}{4th n.n.}  & & & 167 & 119 & & & \\
& \multicolumn{1}{c}{5th n.n.}  & & & 55 & 17 & & & \\ \hline
\label{SK}
\end{largetabular}\\
\begin{center}
\caption{On-site Energies (eV)}
\begin{largetabular}{lccr} 
\hline
& $x$=0.3 & $x$=0.5& $x$=0.7 \\ \hline
xy,x$^2$-y$^2$ & 3.578 & 4.204 & 4.762 \\ 
xz,yz & 2.422 & 3.184 & 3.850 \\ 
z$^2$ & 2.558 & 3.360 & 4.122\\ 
p$_x$,p$_y$ & 0.272 & 0.653 & 1.116 \\ 
p$_z$ & -0.830 & 0.0 & 0.735 \\ \hline
\label{onsite}
\end{largetabular}
\end{center}
\end{table}

        The TB fit yields relative positions of the on-site energy parameters for the Co-d orbitals as expected from simple symmetry
considerations:  the xy,x$^2$-y$^2$ orbitals are highest in energy, the xz,yz orbitals sit lowest and the 3z$^2$-r$^2$ (or a$_{1g}$)
orbital is raised slightly above the latter.  Similarly, the in-plane O onsite energies, p$_x$ and p$_y$, are higher than the
out-of-plane energy, p$_z$. It is important to note that the orbitals familiarly designated as e$_g$' and e$_g$ are both combinations
of the xz,yz and xy,x$^2$-y$^2$ representations in the hexagonal coordinate system.  On a simple triangular lattice (that of the Co
ions alone), these representations do not mix, but the distorted octahedra of O ions breaks local z-reflection symmetry and
facilitates hoppings that would otherwise be forbidden, {\it i.e.} Co-Co hopping between orbitals that are ostensibly orthogonal takes
place through O ions positioned non-symmetrically above and below the Co plane.  A combination of all characters (3z$^2$-r$^2$, xz,yz,
and xy,x$^2$-y$^2$) is present anywhere that a$_{1g}$ and e$_g$' bands mix.  At the very top of the band complex, mostly above the
Fermi energy, the character is purely a$_{1g}$, but elsewhere throughout the energy region e$_{g'}$ character is also present.  Both
band types cross the Fermi level, necessitating a multi-band treatment for accurate reproduction of the DOS and Fermi surfaces.  At
the very least, any model meant to reproduce both a$_{1g}$ and e$_{g'}$ bands on a triangular lattice must include all five Co
d-orbitals, and the symmetry breaking property of the O ions further implies that even this will not necessarily account for all
dispersions.

\begin{figure}
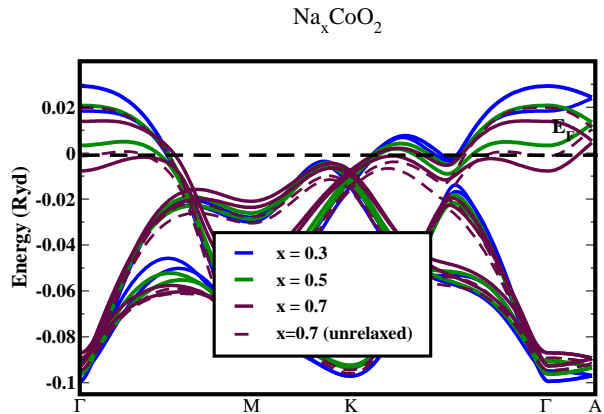

\onefigure[width= 3.1 in]{2.eps}
\caption{The bandstructures of Na$_x$CoO$_2$ for three values of $x$, plotted with Fermi energies aligned. The $x$=0.7 calculation is 
performed at both relaxed and unrelaxed O positions; in the latter the e$_{g'}$ hole surface 
disappears.}
\label{disp}
\end{figure}

We now discuss the main doping and related structural dependencies of the band dispersions.  In the VCA, the addition of Na narrows the
overall bandwidth of what would be, in an ideal octahedral environment, the t$_{2g}$ complex.  The insensitivity of this bandwidth to
c-axis compression suggests that extra charge shifts the balance between competing in-plane interactions, thus decreasing dispersion.  
Hoppings perpendicular to the plane break the degeneracy of the two layers and cause an important splitting of the a$_{1g}$ bands near
the $\Gamma$ point, but have relatively little effect elsewhere.  If inter-layer hopping is eliminated by, for example, a pronounced
expansion of the c-axis length, the six bands collapse onto three doubly degenerate bands and the splitting disappears (see inset of
fig. \ref{fit}).  Not surprisingly, inter-layer coupling increases with growing Na content until at $x$=0.7, the two a$_{1g}$ bands are
separated by as much as 20\% of the total t$_{2g}$ bandwidth along the $\Gamma$-A line.  The final, and in terms of this discussion most
important, effect of Na charge is its determination of the apical O height which in turn establishes the magnitude of the
e$_{g'}$-a$_{1g}$ split.  Coulomb repulsion in the Co-O plane shifts the apical O position upward, easing the octahedral distortion and
lowering the a$_{1g}$ band weight center relative to e$_{g'}$.  Additional charge mainly enters the lowered a$_{1g}$ band, leaving the
e$_{g'}$ hole pockets intact, though diminished in size.  The cumulative result of these effects is summarized in fig. \ref{disp} where
the varying bandwidths and inter-planar splittings corresponding to each Na level are shown.  A second calculation of the $x$=0.7
compound, but with the apical O at an unrelaxed height (we chose the $x$=0.3 height for comparision) illustrates the effect of apical O
re-positioning.  At the unrelaxed (lower) height, only a$_{1g}$ bands cross the Fermi level and e$_{g'}$ states are completely filled,
eliminating the small pockets from the Fermi surface.

\begin{figure}
\twoimages[width= 2.8 in]{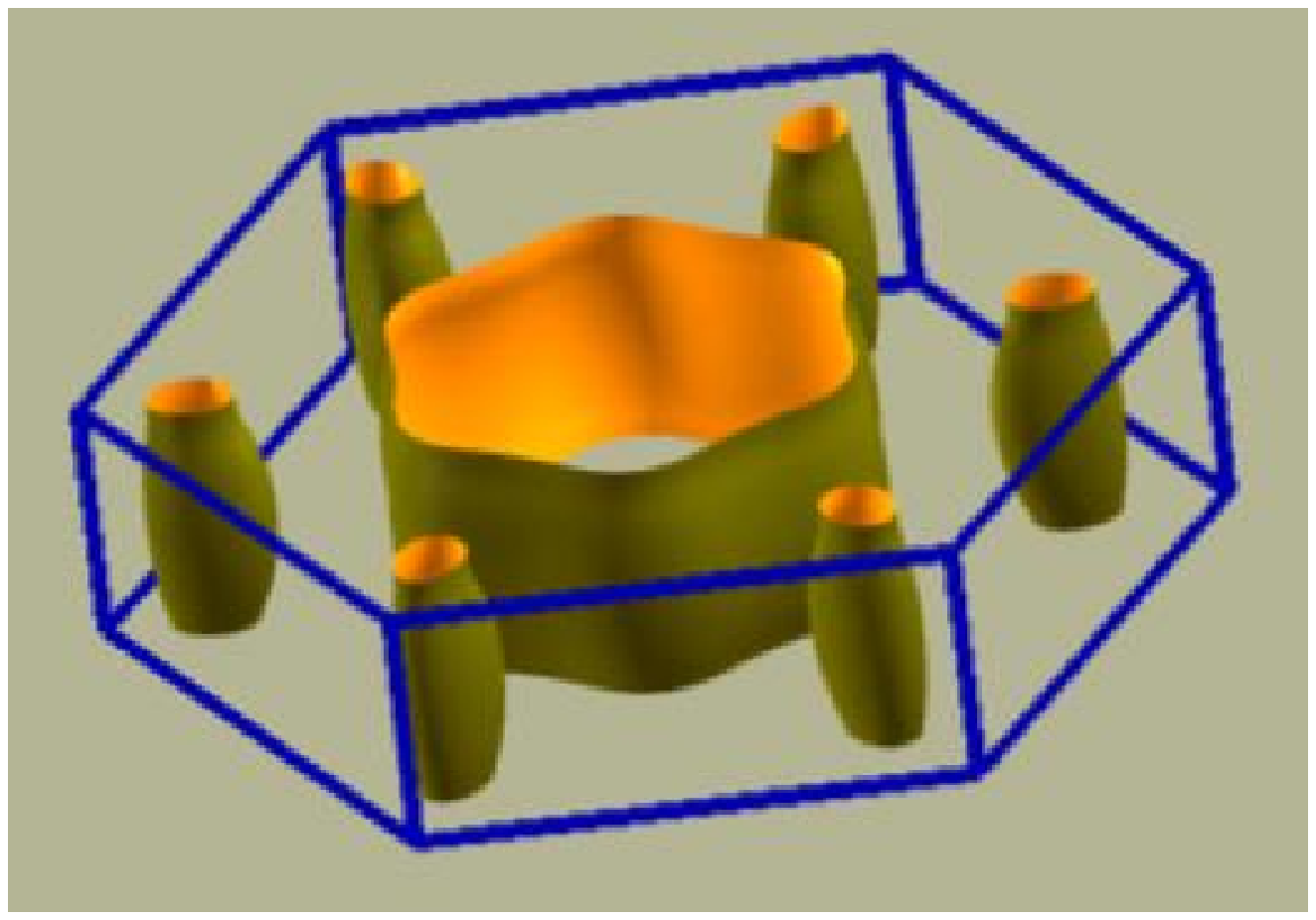}{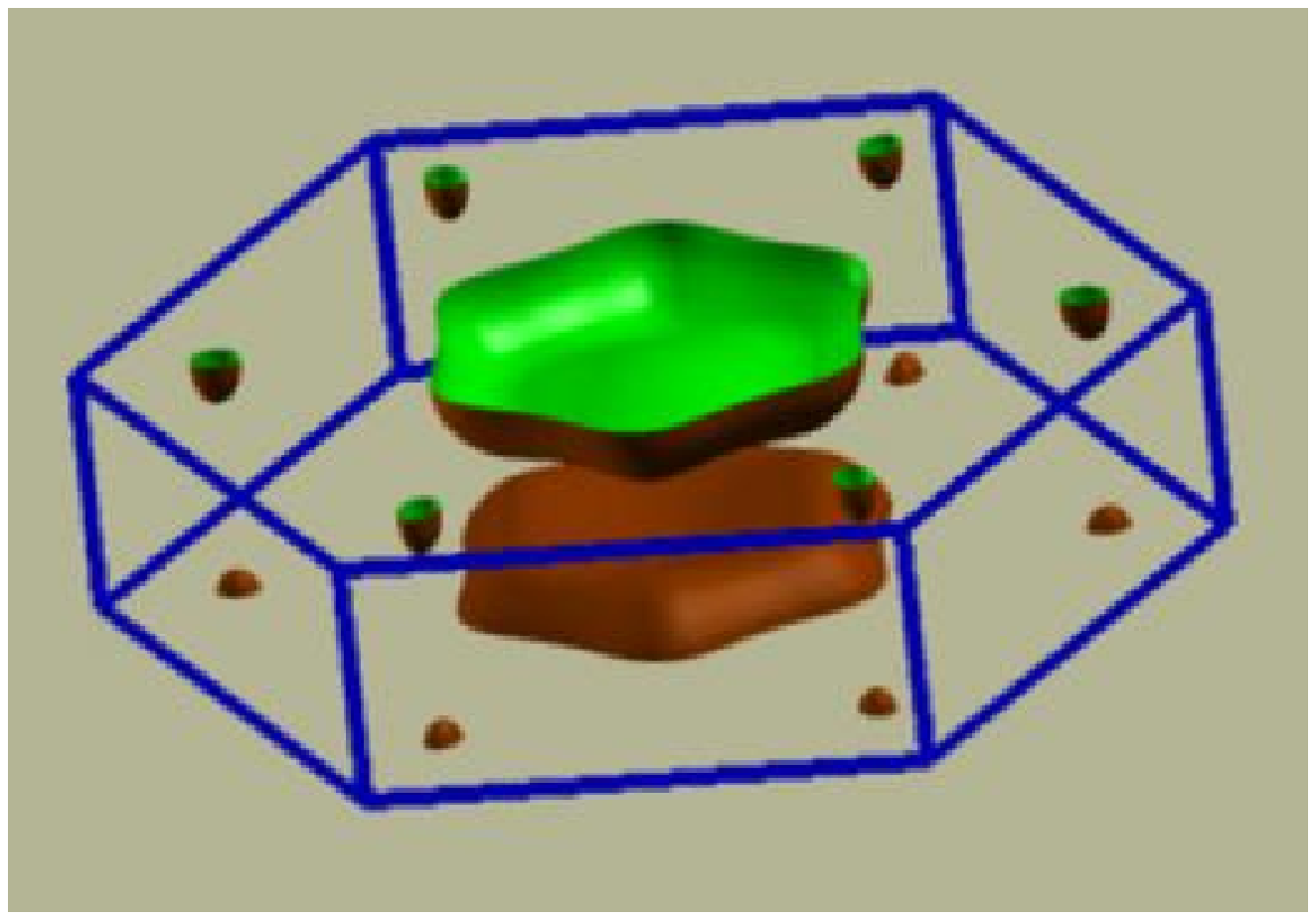} 
\caption{The outer (left) and inner (right) Fermi sheet of Na$_{0.7}$CoO$_2$.  The 
increased charge
present in the Co-O planes both raises the apical oxygen position and increases hybridization, making the 
z-dispersion
non-negligible.  The inner surface shows 3-d dispersion in large (a$_{1g}$) and small (e$_{g'}$) hole pockets} 
\label{fermi}
\end{figure}

We calculated the Fermi surfaces for each doping using our TB parameters and a very dense mesh of nearly 8000 $\vec{k}$-points in the
BZ. Generating the eigenvalues at each of these points in the tight-binding scheme requires less than 1/1000th the amount of computing
time that an LAPW code uses for the same task. For each doping, there is a doubly degenerate a$_{1g}$-like hexagonal large hole pocket
at the top of the BZ, and six doubly degenerate small hole pockets near $K$ which are e$_g$'-like (see fig. \ref{fermi}).  Moving toward
the zone center, the two surfaces disperse away from each other, one surface growing and the other shrinking.  In the $x$=0.3 case, the
dispersion is small for both surfaces, and the system is essentially two-dimensional.  For the $x$=0.5 doping, the dispersion is
stronger, particularly that of the inner or shrinking surface which becomes somewhat hourglass shaped.  At $x$=0.7, the z-dispersion of
the inner surface is so strong that it splits into two separate sheets, disappearing entirely at the zone center.  The outer surface, in
contrast, remains fairly dispersionless.  Both surfaces for this doping are shown in fig. \ref{fermi}.  The increase in z-dispersion
with electron number is due to both the upward shift of the O ions as more charge is dumped into the Co-O plane and to increased
hybridization effects of the charge itself.  When Na is added, holes come mainly out of the shrinking inner surfaces, mainly the
hexagonal one and to a lesser degree the elliptical ones, with the effect that the e$_{g'}$ hole sheets are robust with respect
to doping in the outer surface and are preserved, though barely, in the inner surface.

Recent photoemission experiments \cite{MZH+03,H-Y03} have clearly seen a large Fermi surface consistent with the $a_{1g}$ derived
section predicted in LDA calculations. The small primarily $e_g'$ derived pockets were not seen. This could be due to surface
sensitivity of these states, for example, due to a different surface potential related to surface Na stoichiometry, which would shift
the positions of the these bands or to a relaxation of the O height in the top layer, which would also shift these bands.  As we have
demonstrated, the apical O must be a certain distance from the Co plane for e$_{g'}$ Fermi sheets to exist at the $x$=0.7 doping.  
Therefore, contraction of the Co-O complex at the surface may indeed serve to depress the e$_{g'}$ feature responsible for the small
hole pockets as exemplified by the unrelaxed calculation of Na$_{0.7}$CoO$_2$ in fig. \ref{disp}.  Alternatively, it may be that
correlation effects neglected by the LDA change the position of the e$_{g'}$ band and eliminate the small pockets at dopings where ARPES
measurements have been made, namely at $x$=0.65 and $x$=0.7. K.-W. Lee {\it et al} have suggested, based partly on the same experimental
evidence that prompted division of the phase diagram into three sections, that non-negligable static correlation exists only in the high
Na region \cite{JKK-L04}.  It is possible to produce a relative shift of the $a_g$ and $e_g'$ derived bands in our tight binding
analysis by shifting the corresponding on-site parameters for $x >$ 0.5.  This mimics what would happen in a non-spin-polarized LDA+U
calculation \cite{DAP02} where the tendency toward integer occupation of orbitals strongly disfavors small Fermi surfaces such as the
e$_{g'}$ pockets. We caution that while LDA+U band shifts have been used {\em e.g.} in Ni to improve the agreement between calculated
and experimental Fermi surfaces \cite{IYSYS01}, the physics of applying a static mean field correlation like LDA+U to a metallic system
can be questioned \cite{AGP+03}, as it is known that fluctuation effects as in the dynamical mean field theory (DMFT) are very
important.  A 0.067 Ry shift of the a$_{1g}$ onsite parameter (corresponding to a Hubbard U of 3 eV) eliminates the small ellipses from
the $x$=0.7 Fermi surface and enlarges the hexagonal pocket slightly to accomodate the lost holes. If these hole pockets are indeed
missing due to correlation effects, then the bandstructure can successfully be modelled with a single band (a$_{1g}$) and the
distinction between triangular and pseudo-hexagonal structures discussed previously becomes unimportant.

In conclusion, we have constructed a tight-binding Hamiltonian which can accurately reproduce the LDA Fermi surfaces and band dispersions
of Na$_x$CoO$_2$ for x=0.3, 0.5 and 0.7.  Relaxation of the apical oxygen position at each doping reveals that the Fermi surface is
non-trivially dependent on doping level and that holes preferentially enter and leave from the larger a$_{1g}$-like surface, preserving
the smaller hole pockets as charge is added. Foo {\it et al} \cite{MLF+04} observed three rather distinct regions in the phase space of
$x$, two magnetic and one insulating.  Our calculations show that the system passes from quasi-two-dimensionality at low Na content to
anisotropic three-dimensionality at high Na content.  This may have bearing on the very different magnetic and thermodynamic properties
observed on either side of the insulating phase boundary, though the insulating state itself is not reproduced by the LDA.  We address
the differences in predicted and observed Fermi surfaces and postulate that either surface distortion or possibly correlation could
account for the discrepancies.

\acknowledgements{We would like to thank C. S. Hellberg, M. Z. Hasan, S. Nagler, D. Mandrus and I. Terasaki for contributing useful 
perspective and
discussions.  We are particularly indebted to I. I. Mazin for his insights into symmetry properties and their ramifications.  M.D.J
is supported by a National Research Council associateship.  Research at NRL is funded by the Office of Naval Research.}

\end{document}